\documentclass[journal]{IEEEtran}
\usepackage[cmex10]{amsmath}
\usepackage{amsfonts}
\usepackage{cite}
\usepackage[all]{xy}
\usepackage{booktabs}
\usepackage{url}

\newtheorem{theorem}{Theorem}[section]

\newtheorem{corollary}[theorem]{Corollary}

\newtheorem{lemma}[theorem]{Lemma}
\newtheorem{example}[theorem]{Example}
\def\wt{\mathop{\operatorfont wt}\nolimits}
\newcommand{\ff}{{\mathbb F}}
\newcommand{\ZZ}{{\mathbb Z}}

\def\supp{\mathop{\operatorfont Supp}\nolimits}
\def\wt{\mathop{\operatorfont wt}\nolimits}

\def\Res{\mathop{\operatorfont Res}\nolimits}

\newcommand{\floor}[1]{\left\lfloor#1\right\rfloor}

\newcommand{\setm}[2]{\{\,#1\mid#2\,\}}
\newcommand{\Z}{\ensuremath{\mathbb{Z}}}

\urldef{\duursmaemail}\path|duursma@math.uiuc.edu|
\urldef{\kirovemail}\path|rkirov@ntu.edu.sg|

\begin{document}
\title{Improved Two-Point Codes on Hermitian Curves}
\author{Iwan~Duursma and
        Radoslav~Kirov%
\thanks{Iwan Duursma is with the Department of Mathematics, University of Illinois at Urbana-Champaign, Urbana IL, 61801 USA email: \duursmaemail.}%
\thanks{Radoslav Kirov was with with the Department of Mathematics, University of Illinois at Urbana-Champaign, Urbana IL, 61801. Currently, he is with the School of Physical and Mathematical Sciences, Nanyang Technological University, Singapore 631317, Singapore email: \kirovemail.}}%

\maketitle
\begin{abstract}
One-point codes on the Hermitian curve produce long codes with excellent parameters. 
Feng and Rao introduced a modified construction that improves the parameters while still using one-point divisors. 
A separate improvement of the parameters was introduced by Matthews considering the classical construction but with two-point divisors. 
Those two approaches are combined to describe an elementary construction of two-point improved codes. Upon analysis of their minimum distance and redundancy, it is observed that they improve on the previous constructions for a large range of designed distances. 
\end{abstract}

\section{Introduction}
\IEEEPARstart{H}{ermitian} curves are defined with the equation $y^q+y = x^{q+1}$ over the finite field with $q^2$ elements.
 The number of points is maximal given the genus of the curve and Hermitian codes constructed with Goppa's method have excellent parameters \cite{Tie87, Sti88, YanKum92, KirPel95, Mat01, HomKim05, HomKim06}.
 The curve has as important advantages that the codes are easy to describe and to encode and decode.
 The most studied Hermitian codes are the one-point codes.
 They are obtained by evaluation of functions $f$ in the linear span of $\{x^i y^j : qi+(q+1)j \leq a \}$ for a fixed $a$.
 To a function $f$ corresponds the codeword $(f(P_1),f(P_2),\ldots,f(P_n))$, where the $P_i$, for $i=1,2,\ldots,n$ are distinct rational points on the affine curve.
 Together the codewords generate a code $C(a)$.
\\

Different methods have been presented that give codes with better parameters.
An idea due to Feng and Rao \cite{FenRao95} is to enlarge the code without reducing its distance by carefully selecting extra codewords.
We illustrate their idea for a particular case.
For $q=4$, the codes $C(59)$ and $C(60)$ are of type $[64,54,5]$ and $[64,55,4]$, respectively.
It appears that to increase the dimension of the code we have to accept a smaller distance.
However, it can be shown that words in $C(62) \backslash C(60)$ have weight at least six.
By adding two independent words from $C(62)\backslash C(60)$ to $C(59)$ we obtain a $[64,56,5]$ code.
The parameters of improved one-point Hermitian codes are given in closed form in \cite{BraSul07}.
\\

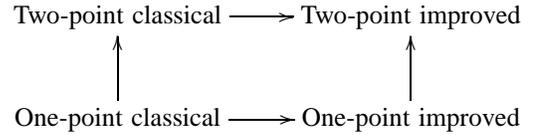
\begin{figure}[ht]
  \centering
\[
\xymatrix{
 \mbox{Two-point classical}\ar[r]  & \mbox{Two-point improved} \\
\mbox{One-point classical} \ar[u] \ar[r] & \ar[u] \mbox{One-point improved}
}
\]
\caption[AG Codes]{Comparison of different constructions for AG codes}
\label{Fig:intro}
\end{figure}

A second idea, first applied by Matthews \cite{Mat01}, is to consider vector spaces of functions that correspond to two-point divisors instead of one-point divisors.
Recently, Homma and Kim gave the complete description of the actual minimum distance for all two-point Hermitian codes \cite{HomKim06}, \cite{Bee07}, \cite{Park10}.
From their description we obtain that the best Hermitian two-point codes are very similar to one-point codes.
They can be defined as subcodes $C'(a)$ of the codes $C(a)$ by removing the functions $1$ and $x$ in the generating set $\{x^i y^j : qi+(q+1)j \leq a \}$ and omitting the point $(0,0)$ in the evaluation, which reduces the code length by one.
This claim is a consequence of the results in \cite{HomKim06,Bee07,Park10} (see Section \ref{S:classical-2pt}).
The codes $C(60)$ and $C(61)$ have subcodes $C'(60)$ and $C'(61)$ of type $[63,53,7]$ and $[63,54,6]$, respectively.
For the two-point codes thus obtained we still have the possibility to improve them further using the Feng and Rao idea.
In this paper, we give bounds for the parameters of Feng-Rao improved two-point codes of special form.
Figure \ref{Fig:intro} summarizes the constructions of Hermitian codes, with arrows pointing in the direction of better codes.
\\


\begin{table}[ht]
  \centering
\begin{tabular}{ccl}
\toprule
Code &~ &Construction \\
\midrule
$[64,54,5]$ &~ &One-point code $C(59)$ \\
$[64,56,5]$ &~ &Improved one-point code $C(59) \oplus \langle c_1, c_2 \rangle$, \\
            &~ &\quad for $c_1, c_2$ such that $C(62) = C(60) \oplus \langle c_1, c_2 \rangle$ \\
$[63,54,6]$ &~ &Two-point code $C'(61)$ \\
\bottomrule 
\end{tabular}
\caption[AG Codes]{Examples of Hermitian codes over $\ff_{16}$ with their construction}
\label{Tbl:intro}\end{table} 


\section{Notation}

We recall the general construction by Goppa of codes from curves. Let $X/\ff$ be an algebraic curve (absolutely irreducible, smooth, projective) of genus $g$ over a finite field $\ff$. Let $\ff(X)$
be the function field of $X/\ff$ and let $\Omega(X)$ be the module of rational differentials of $X/\ff$. Given a divisor $E$ on $X$ defined over $\ff$, let $L(E) = \{ f \in \ff(X) \backslash \{0\} : (f)+E\geq 0 \} \cup \{0\},$ and let $\Omega(E) = \{ \omega \in \Omega(X) \backslash\{0\} : (\omega) \geq E \} \cup \{0\}.$ Let $K$ represent the canonical divisor class. For $n$ distinct rational points $P_1, \ldots, P_n$ on $X$ and for disjoint divisors
$D=P_1+\cdots+P_n$ and $G$, the geometric Goppa codes $C_L(D,G)$ and $C_\Omega(D,G)$ are defined as the images of the maps 
\begin{align*}
\alpha _L :~ &L(G)~\longrightarrow~\ff^{\,n}, ~f \mapsto (f(P_1), \ldots, f(P_n) ). \\
\alpha _\Omega :~ &\Omega(G-D)~\longrightarrow~\ff^{\,n}, 
~\omega \mapsto (\Res_{P_1}(\omega), \ldots, \Res_{P_n}(\omega)).
\end{align*}

The condition that $G$ has support disjoint from $D$ is not essential and can be
removed by modifying the encoding maps $\alpha_L$ and $\alpha_\Omega$ locally at the
coordinates $P \in \supp G \cap \supp D$ \cite{TsfVla07}. We will use both constructions but consider only the case where $X/\ff$ is the
Hermitian curve. The Hermitian curve is the smooth projective curve over $\ff_{q^2}$ with affine equation $y^q + y = x^{q+1}$. It achieves the Hasse-Weil bound with $q^3 + 1$ rational points and genus $g = q(q-1)/2$. For the construction of two-point codes, we fix two distinct rational points $P$ and $Q$. 
The automorphisms of the curve are defined over $\ff_{q^2}$ and form a group of order $q^3(q^3+1)(q^2-1)$. The group acts two-fold transitively on the set of rational points, thus the properties of two-point codes are independent of the choice of $P$ and $Q$. 
The standard choice is to let $P$ be the point at infinity (the common pole of $x$ and $y$)
and $Q$ the origin (the common zero of $x$ and $y$). The equivalent divisors $(q+1)P \sim (q+1)Q$ belong to the hyperplane divisor class $H$.
The divisor sum $R$ of all $q^3+1$ rational points belongs to the divisor class $(q^2-q+1)H$ and the canonical divisor class $K=(q-2)H$ \cite{Tie87,Sti88,TsfVla07}.
\\

Let $\{G_i\}$ be an increasing sequence of divisors, where $G_i-G_{i-1}$ is a rational point. From each divisor we can produce a code by using either construction method. If the divisor sequence is long enough we produce a sequence of nested codes, containing a code of each possible dimension. The sequences that we are primarily interested in are the sequence of Hermitian one-point codes $0 \subseteq \cdots \subseteq C(a-1) \subseteq C(a) \subseteq \cdots \subseteq \ff^n$, for $C(a) = C_L(D,aP)$, $D=R-P$, and the special sequence of Hermitian two-point codes 
$0 \subseteq \cdots \subseteq C'(a-1) \subseteq C'(a) \subseteq \cdots \subseteq \ff^n$, for $C'(a) = C_L(D,aP-2Q)$, $D=R-P-Q$. 
The choice of the second sequence is motivated by the observation that the codes $C_L(D,aP-2Q)$ have optimal minimum distance among all two-point codes $C_L(D,mP+nQ)$ with $m+n=a-2$ (see Section \ref{S:classical-2pt}).
It shows that the classical two-point codes along that sequence have best parameters among classical two-point codes. 
For a sequence of nested codes, the minimum distance can be estimated with the Feng-Rao bound \cite{FenRao93,Duu93,KirPel95, HoeLinPel98}, or one of its generalizations \cite{Bee07,AndGei08, DuuKir09, DuuPar10, DuuKirPar10}.  
The Feng-Rao bound uses estimates for the weight of codewords that belong to one of the codes in the sequence but not to the immediate subcode. In our case, we use lower bounds for the weight of words in $C(a) \backslash C(a-1)$ or $C'(a) \backslash C'(a-1)$. It is often the case that the sequence of coset minimum weights is not monotone. By the classical construction to obtain a code with designed distance $\delta$ one chooses the largest code $C(a)$ in the sequence just before the weights in
$C(a+1) \backslash C(a)$ fall below $\delta$, regardless of the weights after that in the differences $C(a+2) \backslash C(a+1)$, etc.. Feng and Rao \cite{FenRao95} observed that those cosets can be used if we modify the construction. For a fixed designed distance $\delta$, use as generators for the Feng-Rao improved code one word from each difference $C(a) \backslash C(a-1)$ for which the minimum weight is greater than or equal to $\delta$. It is 
convenient to count the number of cosets that need to be removed (i.e. cosets with weight less than $\delta$). The number $r=r(\delta)$ of such cosets
gives the redundancy (i.e. the dimension of the dual code) as a function of the designed distance. \\

For a given divisor $D=P_1+\ldots+P_n$, let $\{G_i\}$ be a long enough sequence of divisors $\{G_i\}$ (i.e. producing codes
$C_L(D,G_i)$ and $C_\Omega(D,G_i)$ of every possible dimension). The improved codes with designed distance $\delta$ constructed with the sequence 
have parameters $[\deg(D),\deg(D)-r(\delta),\geq \delta]$, where 
\begin{align*}
r(\delta) = 
\;&|\{i: 0 < \min \wt (C_L(D,G_i)\backslash C_L(D,G_{i-1})) <\delta\}|,\\
& ~~ \text{for the sequence of codes $\{ C_L(D,G_i) \}$,} \\
r(\delta) = 
\;&|\{i: 0 < \min \wt (C_\Omega(D,G_i)\backslash C_\Omega(D,G_{i+1})) <\delta\}|, \\
&~~\text{for the sequence of codes $\{ C_\Omega(D,G_i) \}$.}
\end{align*}

Here we use the convention that $\min \wt (\emptyset) = 0$, so that $\min \wt \; C_L(D,G_i)\backslash C_L(D,G_{i-1}) = 0$ for $C_L(D,G_i) = C_L(D,G_{i-1}).$

\section{Improved codes along two sequences}

The one-point Feng-Rao methods have been generalized to give coset bounds for two-point divisors \cite{Bee07 ,DuuPar10,DuuKir09,DuuKirPar10}. Coset bounds for Hermitian two-point codes were first obtained in \cite[Fact 13, Proposition 14]{Bee07} and \cite{SKP07}.
We use the formulation in \cite[Proposition 3.1]{Park10}.

\begin{theorem}[\cite{Bee07, SKP07}; \cite{Park10}] \label{T:coset}
Let $C=dH-aP-bQ$, where $0\leq a,b \leq q$ and $P \notin \supp D$. Then 

\begin{multline*}
\min \wt C_\Omega(D,K+C) \backslash C_\Omega(D,K+C+P) \geq \\
\left\{
     \begin{array}{ll}
       \deg(C), & \mbox{ for } a-d < 0. \\
       a(q-1-a+d)\\
       +\max\{0,a-b\}, & \mbox{ for } 0 \leq a-d \leq q-1. \\         
        0, & \mbox{ for } a-d > q-1. \\
     \end{array}
   \right. 
\end{multline*}
\end{theorem}

To apply the improved Feng-Rao construction to two-point codes we need to select a particular sequence of divisors (in the one-point case there is no choice to be made). In exhaustive computations over $\ff_{2^{2k}}$, $k=1,2,3,4,5,$ and $\ff_{3^{2k}}$, $k=1,2,3,$ we observed that among all possible sequences, the sequence $\{ iP+Q \}$ produced the optimal redundancy $r(\delta)$ for any designed distance $\delta$. In the rest of this paper we therefore study the parameters of the following two sequences:

\begin{enumerate}
\item One-point codes $C_\Omega(D,G_i)$, for $D=R-P$ and $G_i=iP.$
\item Two-point codes $C_\Omega(D,G'_i)$, for $D=R-P-Q$ and $G_i'=iP+Q$.
\end{enumerate}

Using $K = (q-2)H$, the previous theorem immediately gives bounds for the minimum weights of cosets in each of the sequences (1) and (2).

\begin{corollary}\label{C:simple}
  Write $iP=(d+q-2)H-aP$, for unique $d \in \ZZ$ and $0\leq a \leq q$. If $\supp D \cap \{P\}=\emptyset$, then \\
\begin{multline*}
(1) \min \wt C_\Omega(D,iP) \backslash C_\Omega(D,(i+1)P) \geq \\
\left\{
     \begin{array}{ll}
       (q+1)d-a, & \mbox{ for } a-d < 0. \\
       a(q-a+d), & \mbox{ for } 0 \leq a-d \leq q-1. \\         
        0, & \mbox{ for } a-d > q-1. \\
     \end{array}
   \right. 
 \end{multline*}
 \begin{multline*}
(2) \min \wt C_\Omega(D,iP+Q) \backslash C_\Omega(D,(i+1)P+Q) \geq 
\\ \left\{
     \begin{array}{ll}
       (q+1)d-a+1, & \mbox{ for } a-d-1 < 0. \\
       a(q-a+d), & \mbox{ for } 0 \leq a-d-1 \leq q-1. \\         
        0, & \mbox{ for } a-d-1 > q-1. \\
     \end{array}
   \right.
\end{multline*}
\end{corollary}

Table \ref{Tab:inconsistent} illustrates the lower bounds in Corollary \ref{C:simple} for the Hermitian curve over $\ff_{16}$ (the case $q=4$). The lower bounds are given for every inclusion in $\ff^n = C_\Omega(D,-P) \supseteq C_\Omega(D,0) \supseteq \cdots \supseteq C_\Omega(D,23P)$ (the row $G_i$) as well as 
$\ff^n = C_\Omega(D,Q-P) \supseteq C_\Omega(D,Q) \supseteq \cdots \supseteq C_\Omega(D,Q+23P)$ (the row $G'_i$).
For $i=-1$, $C_\Omega(D,-P) \backslash C_\Omega(D,0)$ contains all words that are not orthogonal to the all-one word, and the minimum weight is $1$. 
For $iP \geq 2K+2P$ (or $i \geq 2(q^2-q-1)=22$), the coset bounds for $C_\Omega(D,iP) \backslash C_\Omega(D,(i+1)P)$ agree with the Goppa lower bound 
$i-(2g-2)$ for the minimum distance of $C_\Omega(D,iP).$  
\begin{table*}[ht]
  \centering
\begin{tabular}{@{}cccccccccccccccccccccccccc@{}}
\toprule
$i$   & -1 & 0& 1& 2& 3& 4& 5& 6& 7& 8& 9& 10& 11& 12& 13& 14& 15& 16& 17& 18& 19& 20& 21 & 22 \\ 
\midrule
$d$  & -2 & -2 & -1 & -1 & -1 & -1 & -1 & 0 & 0 & 0 & 0 & 0 & 1 & 1 & 1 & 1 & 1 & 2 & 2 & 2 & 2 & 2 & 3 & 3 \\ 
$a$  & 1 & 0 & 4 & 3 & 2 & 1 & 0 & 4 & 3 & 2 & 1 & 0 & 4 & 3 & 2 & 1 & 0 & 4 & 3 & 2 & 1 & 0 & 4 & 3 \\ 
\midrule
$G_i$ & 1& 0& 0& 0& 2& 2& 0& 0& 3& 4& 3& 0& 4& 6& 6& 4& 5& 8& 9& 8& 9& 10& 12& 12 \\
$G'_i$ & 1& 0& 0& 0& 2& 2& 0& 0& 3& 4& 3& 1& 4& 6& 6& 5& 6& 8& 9& 9& 10& 11& 12& 13 \\ 
\bottomrule 
\end{tabular}
\caption{Bounds on coset minimum weights using Corollary \ref{C:simple}}
\label{Tab:inconsistent}
\end{table*} 

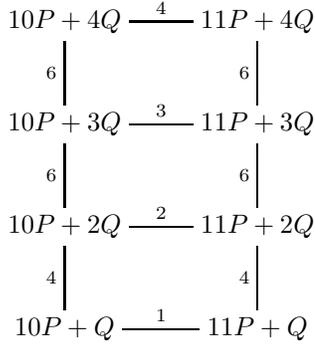
\begin{figure}
\[
\xymatrix{
10P+4Q \ar@{-}[r]^4 & 11P+4Q \\
10P+3Q \ar@{-}[u]^6 \ar@{-}[r]^3 & 11P+3Q \ar@{-}[u]^6  \\
10P+2Q \ar@{-}[u]^6 \ar@{-}[r]^2 & 11P+2Q \ar@{-}[u]^6  \\
10P+Q \ar@{-}[u]^4 \ar@{-}[r]^1 & 11P+Q \ar@{-}[u]^4  
}
\]
\caption{Illustration of an inconsistency in the coset bounds coming from Theorem \ref{T:coset}}
\label{Fig:inconsist}
\end{figure}

More advanced methods \cite{DuuKirPar10} do not improve the estimates for the sequence $\{ G_i \}$. Meanwhile some estimates for $G'_i$ can be improved with a simple observation illustrated in Figure \ref{Fig:inconsist}. In the figure we see a grid of divisors over the Hermitian curve with $q=4$ with edge labels giving the bound on minimum coset weights coming from Theorem \ref{T:coset}.

Starting at $10P+Q$ the path going first up and then across guarantees $\min \wt C_\Omega(D,10P+Q) \backslash C_\Omega(D,11P+4Q)$ is at least $4$. Since the path going across first and then up also measures the same quantity, all estimates along that path are at least $4$. Performing such improvements in the general case leads to the following theorem.

\begin{theorem}\label{T:impcoset}
Let $iP=(d+q-2)H-aP$, for unique $d \in \ZZ$ and $0\leq a \leq q$. If $\supp D\cap\{P,Q\}=\emptyset$, then
\begin{multline*}
\min \wt C_\Omega(D,iP+Q) \backslash C_\Omega(D,(i+1)P+Q) \geq \\
 \left\{
     \begin{array}{lr}
        (q+1)d-a+1 & \mbox{ for } d > q-1 \\
       qd+q-a & \mbox{ for } a \leq d \leq q-1  \\
       a(q-a+d) & \mbox{ for } 0 \leq a-d-1 \leq q-1 \\         
        0 & \mbox{ for } a-d-1 > q-1 \\
     \end{array}
   \right. 
\end{multline*}
\end{theorem}

\begin{IEEEproof}
We only improve the previous estimate in the case $a \leq d \leq q-1$. We use divisor paths shown in Figure \ref{Fig:impcoset}; where $iQ$ is rewritten as $H-(q+1-i)Q$.

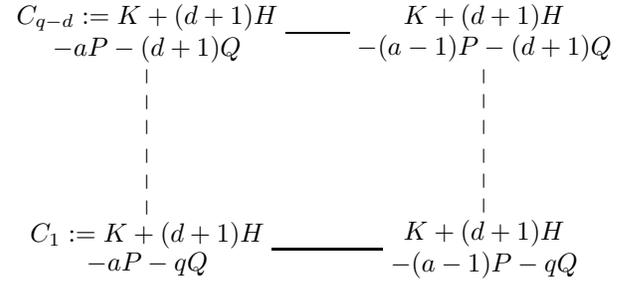
\begin{figure}[ht]
\[
\xymatrix{
\txt{$C_{q-d}:=K+(d+1)H$\\$-aP-(d+1)Q$} \ar@{-}[r] & \txt{$K+(d+1)H$\\$-(a-1)P-(d+1)Q$} \\
\ar@{--}[u]&\ar@{--}[u]\\
\txt{$C_1:=K+(d+1)H$\\$-aP-qQ$} \ar@{--}[u] \ar@{-}[r] & \txt{$K+(d+1)H$\\$-(a-1)P-qQ$} \ar@{--}[u]  
}
\]
\caption{Illustation of the proof of Theorem \ref{T:impcoset}}
\label{Fig:impcoset}
\end{figure}

Note that $d \leq q-1$ is required to ensure that the orientation in the diagram is correct. Using the assumptions and Theorem \ref{T:coset} the bound on the $P$-coset at $C_{q-d}$ equals $\deg(C_{q-d})= qd+q-a$. For the remaining $Q$-cosets at $C_j$ we use Theorem \ref{T:coset} but with $P$ and $Q$ swapped. The divisors $C_j$ equal $K+(d+1)H-aP-jQ$ for the range $j \in \{d+2,\ldots,q\}$. Using the assumptions the bound simplifies to $j(q-i+d)+j-a$. As a function of $j$ its minimum is at $j=q$ and equals exactly $qd+q-a$.

Thus the path consisting of first adding $q-d-1$ times $Q$ and then $P$ has minimum weight at least $qd+q-a$, so every edge of any other path to the same divisor has at least that weight. In particular the $P$-coset at $K+(d+1)H-aP-qQ = iP+Q$  
\end{IEEEproof}

Note that in the proof above we used Theorem \ref{T:coset} with both $P$ and $Q$. 
As a consequence we need to remove both from the support of $D$. 
The improved two-point codes are one coordinate shorter than the improved one-point codes. 
Since the action of the automorphism group of the curve on the set of rational points is two-fold transitive, the automorphism group of a one-point code acts transitively on the set of coordinates.
Thus the minimum distance of the code is preserved under shortening. 
This feature makes it easy to compare two-point codes with similar one-point codes of different length. \\

Going back to our running example with $q=4$, which is the Hermitian curve over $\ff_{16}$, we find six improvements for the coset bounds in the
sequence $\{ G'_i \}$.
\begin{table*}[ht]
  \centering
\begin{tabular}{@{}cccccccccccccccccccccccccc@{}}
\toprule
$i$   & -1 & 0& 1& 2& 3& 4& 5& 6& 7& 8& 9& 10& 11& 12& 13& 14& 15& 16& 17& 18& 19& 20& 21 & 22 \\ 
\midrule
$d$  & -2 & -2 & -1 & -1 & -1 & -1 & -1 & 0 & 0 & 0 & 0 & 0 & 1 & 1 & 1 & 1 & 1 & 2 & 2 & 2 & 2 & 2 & 3 & 3 \\ 
$a$  & 1 & 0 & 4 & 3 & 2 & 1 & 0 & 4 & 3 & 2 & 1 & 0 & 4 & 3 & 2 & 1 & 0 & 4 & 3 & 2 & 1 & 0 & 4 & 3 \\ 
\midrule
$G_i$ & 1& 0& 0& 0& 2& 2& 0& 0& 3& 4& 3& 0& 4& 6& 6& 4& 5& 8& 9& 8& 9& 10& 12& 12 \\ 
$G'_i$ & 1& 0& 0& 0& 2& 2& 0& 0& 3& 4& 3& {\bf 4}& 4& 6& 6& {\bf 7}& {\bf 8}& 8& 9& {\bf 10}& {\bf 11}& {\bf 12}& 12& 13 \\
\bottomrule 
\end{tabular}
\caption{Bounds on coset minimum weights using Theorem \ref{T:impcoset}}
\end{table*} 

Computing the redundancy along each sequence amounts to counting the coset bounds that are strictly between $0$ and the designed distance $\delta$. 
For the one-point sequence $G_i$ the redundancy is available in a closed form in the work of Bras-Amor\'{o}s-O'Sullivan \cite{BraSul07}. 
We will not present it here as it is rather long. 
Instead we present the improvement of the redundancies of improved two-point codes over improved one-point codes for arbitrary designed distance $\delta$.\\
\begin{corollary}
  \label{C:diff}
Let $q \leq \delta < q^2$ and $\delta = dq + b$ for unique $0 < d \leq q-1$ and $0 < b \leq q$, then 
\[
r_{G_i}(\delta) - r_{G'_i}(\delta) = \left\{
\begin{array}{lr}
  b-1 & \text{ if } b \leq d \leq q-b.\\
  q-d-1 & \text{ if } b \leq d \text{ and } q-b < d.\\
  q-b & \text{ if } q-b < d < b. \\
  d & \text{ if } d < b  \text{ and } d \leq q-b .\\
\end{array}
\right.
\]
\end{corollary}
\begin{IEEEproof}
  For a fixed designed distance $\delta$ we improve the redundancy by one when the $G_i$ coset bound is below $\delta$ while the $G'_i$ coset bound is above or equal. 
  Comparing Corollary \ref{C:simple} and Theorem \ref{T:impcoset} we observe that the coset bounds differ if $a\leq d \leq q$ and if $d > q$. 
  Because $\delta \leq q^2$ we can ignore the second case. 
  The number of such improvements is 
  \[
   |\{ (a,d') : 0 \leq a \leq d' \leq q, (q+1)d' - a < \delta \leq (d'+1)q-a\}|.
  \]
  Combining the inequalities we see that the only admissible $d'$ is $d$. Thus, 
\begin{multline*}
 r_{G_i}(\delta) - r_{G'_i}(\delta)\\
 = |\setm{a \in \Z}{0 \leq a \leq d \text{ and } d - b < a \leq  q - b}| - 1.
\end{multline*}
We subtract one for each case when $G_i$ is zero, while $G'_i$ is not.
This only occurs for $a=d=0$ and the $G'_i$ coset bound is $q$ in this case.
Since $q < \delta$, for all $\delta$ we subtract one.
The final result is a case analysis of the quantity.
\end{IEEEproof}
\begin{example}
  \label{Ex:goodfamily}
  Let $q$ be even and $\delta = q(q+1)/2$.
  Then $d=b=q/2$ and the two-point improved code along $G'_i$ has redundancy $q/2 - 1$ less than the one-point improved code.
  This is the maximum possible difference between a two-point and a one-point improved code.
\end{example}

The bounds in this section were formulated for residue codes $C_\Omega(D,G).$ To apply the bounds to
evaluation codes, we include the following lemma. For the Hermitian curve, let $H=(q+1)P$, let
$R$ denote the divisor sum of all $q^3+1$ rational 
points and let $K$ denote the canonical divisor class. 

\begin{lemma} \label{L:evres}
For the Hermitian curve, each of the following choices of $D, G^\ast$ and $G$ describes
a pair of equivalent codes $C_\Omega(D,G^\ast)$ and $C_L(D,G)$. 
\begin{align*}
(1)~  &D = R-P & &G^\ast = K+dH-aP,  \\
     &        &   &G = R-dH+aP-P. \\
(2)~  &D=R-P-Q & &G^\ast = K+dH-aP+Q, \\
     &        &   &G = R-dH+aP-P-2Q. \\
(2')~ &D=R-P-Q & &G^\ast = K+dH-aP-qQ, \\
     &        &   &G = R-(d-1)H+aP-P-2Q.
\end{align*}
\end{lemma}

\begin{IEEEproof}
In general, codes $C_L(D,G)$ and $C_\Omega(D,G^\ast)$ are equivalent provided that 
$G+G^\ast \sim K+D$ \cite{Sti09}.
\end{IEEEproof}

The divisor sum $R$ of all $q^3+1$ rational points belongs to the divisor class 
$(q^2-q+1)H=(q^3+1)P$ and the
evaluation codes $C_L(D,G)$ in the lemma are one-point codes (case 1) and special
cases of two-point codes (cases 2, 2'). It can be shown that, for the Hermitian codes in the lemma,
the equivalences are equalities. 
The argument for one-point codes is based on exhibiting a differential that has a simple pole with residue $1$ for every point in $D$ \cite{Sti88, MunSepTor09}. 
The same argument along with the same choice of differential works for two-point codes.

Using Lemma \ref{L:evres} we restate Corollary \ref{C:simple} and Theorem \ref{T:impcoset} for evaluation codes. 
\begin{corollary}
Let $iP=(d+q-2)H-aP$, for unique $d \in \ZZ$ and $0\leq a \leq q$ and let $G = (q^2-1)H-iP-P$.
If $D = R - P$, then
\begin{multline*}
  \min \wt C_L(D,G) \backslash C_L(D,G-P) \geq \\
  \left\{
     \begin{array}{ll}
       (q+1)d-a, & \mbox{ for } a-d < 0. \\
       a(q-a+d), & \mbox{ for } 0 \leq a-d \leq q-1. \\         
        0, & \mbox{ for } a-d > q-1. \\
     \end{array}
   \right.
\end{multline*}
If $D = R - P - Q$, then
\begin{multline*}
  \min \wt C_L(D,G+Q) \backslash C_L(D,G+Q-P) \geq\\
  \left\{
     \begin{array}{lr}
        (q+1)d-a+1 & \mbox{ for } d > q-1 \\
       qd+q-a & \mbox{ for } a \leq d \leq q-1  \\
       a(q-a+d) & \mbox{ for } 0 \leq a-d-1 \leq q-1 \\         
        0 & \mbox{ for } a-d-1 > q-1 \\
     \end{array}
   \right. 
\end{multline*}
\end{corollary}
\section{Classical Hermitian two-point codes} \label{S:classical-2pt} 
Classical Hermitian two-point codes are evaluation codes $C_L(D,G)$ with Goppa divisor $G=mP+nQ$. The actual minimum distance for Hermitian two-point codes was determined by Homma and Kim \cite[Theorem 5.2, Theorem 6.1]{HomKim05} (cases $n=0$ and $n=q$), \cite[Theorem 1.3, Theorem 1.4]{HomKim06} (cases $0 < n < q$). Using order bound techniques, Beelen \cite[Theorem 17]{Bee07} gives lower bounds for the cases 
$\deg G > \deg K$ (i.e. $m+n > (q-2)(q+1)$), and Park \cite[Theorem 3.3, Theorem 3.5]{Park10} 
for all cases. Park moreover shows that the lower bounds are sharp and that they correspond
to the actual minimum distance. We recall these results and derive from it that among all divisors $G=mP+nQ$ of given degree, the optimal minimum distance is attained for a choice of the form $G=aP-2Q.$
%
%

\begin{theorem}[\cite{HomKim05, HomKim06}; \cite{Bee07,Park10}]
\label{T:dist} Let $C$ be a divisor such that $C\neq 0$, $\deg C \geq 0$ and $C = dH -aP -bQ,$ for $d \geq 0$ and for $0 \leq a,b \leq q.$ For $G=K+C$, the algebraic geometric code $C_\Omega(D,G)$
has minimum distance $\delta$, where \\
   Case 1: If $a,b \leq d$
  \[
\delta = \deg C.  
  \]
   Case 2a: If $b \leq d \leq a$ and $\supp D\cap\{P\} = \emptyset $   
    \[
\delta = \deg C + a-d. 
\]
   Case 2b: If $a \leq d \leq b$ and $\supp D\cap\{Q\} = \emptyset $   
    \[
\delta = \deg C + b-d.
\]
   Case 3a: If $d \leq a \leq b, a < q$ and $\supp D\cap\{P,Q\} = \emptyset $   
    \[
\delta = \deg C + a-d + b-d.
\]
   Case 3b: If $d \leq b \leq a, b < q$ and $\supp D\cap\{P,Q\} = \emptyset $   
    \[
\delta = \deg C + a-d + b-d.
\]
   Case 4: If $d \leq b = a = q$ and $\supp D\cap\{P,Q\} \neq \{P,Q\}$   
    \[
\delta = \deg C + q-d.
\]

\end{theorem}
\begin{IEEEproof}
The results were first obtained in \cite{HomKim05, HomKim06}. In the approach used in \cite{Bee07, Park10}, lower bounds for $\delta$ follow from 
repeated application of Theorem \ref{T:coset} for P-cosets or Q-cosets. For a proof that those lower 
bounds are sharp see \cite{Park10}.
\end{IEEEproof}

We isolate the cases that attain the best possible minimum distance for a given degree of the
divisor $G$.

\begin{corollary} \label{C:best}
Let $C$ be a divisor such that $C \neq 0$, $\deg C \geq 0$ and $C = dH -aP -bQ,$ for $d \geq 0$ and for $0 \leq a,b \leq q,$ and let $G=K+C.$ For a given degree
of the divisor $G$, the optimal minimum distance for a two-point $C_\Omega(D,G)$ is attained for $b=q$
and it equals\\
   Case 1: If $q \leq d$
  \[
\delta = \deg C.  
  \]
   Case 2: If $a \leq d \leq q$ and $\supp D\cap\{Q\} = \emptyset $   
    \[
\delta = \deg C + q - d.
\]
   Case 3: If $d \leq a < q$ and $\supp D\cap\{P,Q\} = \emptyset $   
    \[
\delta = \deg C + a-d + q-d.
\]
   Case 4: If $d \leq a = q$ and $\supp D\cap\{P,Q\} \neq \{P,Q\}$   
    \[
\delta = \deg C + q - d.
\]

\end{corollary}

\begin{IEEEproof}
Cases $1),2)$ and $3)$ of Theorem \ref{T:dist} can be combined as $\deg C + \max\{0,a-d\} + \max\{0,b-d\}$. Assume without loss of generality that $ a \leq b \leq q-1$. Then we can increase $b$ by one and decrease $a$ by one to produce a divisor of same degree. There are two possibilities. If $a \neq 0$, any $\max\{0,a-d\}$ decrease would be offset by an increase in $\max\{0,b-d\}$. If $a=0$, the new divisor will have $d'=d+1$, $a'=q$, and $b'=b+1$. Thus $\max\{0,b-d\}$ is unchanged and
$\max\{0,a-d\}=0$ becomes $\max\{0,a'-d'\}=\max\{0,q-1-d\}$, with a strict increase if $d < q-1$ and no change otherwise. 
\end{IEEEproof}

We use Lemma \ref{L:evres} to reformulate the optimal cases as evaluation codes.


\begin{theorem} 
\label{T:best2ptev}
Let $C = dH-aP-qQ$, for $d \in \ZZ,$ and for $0 \leq a,b \leq q,$ and let $m = q^3 + 1 - \deg C$. 
\[
\begin{array}{lcl}
\text{\rm{Case 1 : }} a \leq d \mbox{ or } d \leq a = q) \\
\qquad d(C_L(R-Q,(m+1)P-2Q)) = \deg C + q - d \\
\text{\rm{Case 2 : }} d \leq a < q\\
\qquad d(C_L(R-P-Q,mP-2Q))= \deg C + q + a - 2d \\
\end{array}
\]
In each case, the code has the highest minimum distance among all two-point codes $C_L(D,G)$ with
$G$ of the same degree.
\end{theorem}

Knowing that the codes in the sequence $\{ C_L(D,G=mP-2Q) \}$ are optimal when $\deg G \geq \deg K$ means that improved two-point codes obtained with the sequence are at least as good
as any classical two-point code. 
\begin{corollary}
  The two point improved codes along $G_i'=iP+Q$ strictly improve on the best two-point classical code for a designed distance $\delta \in [q,(q-1)(q-2\sqrt{q-1})]$.  
  Moreover, in the range $q < \delta \leq q^2$ the ratio of two-point improved codes which have strictly lower redundancy than both two point classical codes and one-point improved codes is at least
  \[
  1 -\frac{4\sqrt{q-1}+4}{q}.  
  \]
  \begin{IEEEproof}
    By Corollary \ref{C:best} the best two-point classical codes have $G=iP+Q$, thus the improved codes are at least as good. 
    Strict improvements occur when the coset bounds given in Theorem \ref{T:impcoset} decrease for increasing degree and the designed distance exceeds the smaller but not the larger of those two bounds. 
    For a fixed $d$ for $a$ in the range $[d+1,(g+q)/2]$, this condition is satisfied.
    Thus, for 
  \[
  \delta \in \left[ (q-1)(d+1)+1,\dots, \left(\frac{d+q}{2}\right)^2 \right] \]
  such that\[ \qquad 0 \leq d < q \] 
  there is a strict improvement.  The sets are overlapping for $0\leq d \leq q-2-2\sqrt{q-1}$, thus obtaining a bound on the difference with two-point classical codes.
  By Corollary \ref{C:diff} there are $3q-4$ designed distances between $q$ and $q^2$ for which two-point improved codes have the same redundancy as one-point improved codes.
  We obtain the final result by assuming conservatively that the cases where two-point improved codes have the same redundancy as one-point improved codes do not overlap the cases where two-point improved codes have the same redundancy as two-point classical codes.
\end{IEEEproof}
\end{corollary}
\begin{example}
  Continuing Example \ref{Ex:goodfamily} we examine designed distance $\delta = q(q+1)/2$ for even $q$.
  The two-point classical code for $C = q/2H - q/2P + Q = (q/2+1)H - q/2P - qP$ meets $\delta$ by Corollary \ref{C:best}.
  Examining the cosets with $d=q/2$ and $q/2 < a \leq q$ using Theorem \ref{T:impcoset}, we observe that the cosets with minimum weight equal or over $\delta$ occur for 
  \[a \in \left[\frac{3}{4}q - \alpha, \frac{3}{4}q + \alpha\right]\qquad \text{where} \qquad \alpha = \floor{\frac{\sqrt{q(q-8)}}{4}}.\]
  Thus $C$ is the largest two-point classical for code with distance $\delta$, and for $q \geq 8$ the two-point improved code along $G'_i$ has strictly higher redundancy.
  The redundancy difference is at least $2\alpha $.
\end{example}
Along with Example \ref{Ex:goodfamily} this shows that for any even $q$ the two-point improved code with $\delta = q(q+1)/2$ has strictly lower redundancy than the best two-point classical code and the best one-point improved code for the same distance.
In both cases the gain in the redundancy is $O(q)$.
\section{Experimental Results}

Tables \ref{Tab:16} and \ref{Tab:64} compare the redundancies of one- and two- point codes (both classical and improved) for Hermitian curves over $\ff_{16}$ and $\ff_{64}$. Since we have not theoretically established that a choice $G'_i=iP+Q$ produces the best improved two-point codes, for the table entries we checked all possible sequences.
We see that for the Hermitian curve over $\ff_{16}$ the parameters of two-point improved codes are matched by either a one-point improved code or a two-point classical code. 
However, for the Hermitian curve over $\ff_{64}$, there exist designed distances for which two-point improved codes have strictly lower redundancy than both one-point improved codes and two-point classical codes.

\begin{table}\centering
\begin{tabular}{@{}rcrrcrrcr@{}}\toprule
$\delta\backslash r$ &\phantom{a}& \multicolumn{2}{c}{One-point} &\phantom{a}& \multicolumn{2}{c}{Two-point} &\phantom{a}&  \\
\cmidrule{3-4} \cmidrule{6-7}
&& C & I && C & I &&\\
\midrule
3 && 3 & 3 && 3 & 3 && 0\\
4 && 6 & 5 && 6 & 5 && 0\\
5 && 10 & 8 && 8 & 8 && 0\\
6 && 11 & 9 && 8 & 8 && 0\\
7 && 11 & 11 && 10 & 10 && 0\\
8 && 11 & 11 && 11 & 11 && 0\\
9 && 14 & 13 && 13 & 13 && 0\\
10 && 15 & 15 && 14 & 14 && 0\\
11 && 16 & 16 && 15 & 15 && 0 \\
\bottomrule
\end{tabular}
\caption{Hermitian Curve over $\ff_{16}$. Optimal redundancy $r$ for given minimum distance $\delta$. The last column measures the improvement of the two-point improved code over the lowest redundancy in the previous columns.}    
\label{Tab:16}
\end{table}

\begin{table}\centering
\begin{tabular}{@{}rcrrcrrcr@{}}\toprule
$\delta\backslash r$ &\phantom{a}& \multicolumn{2}{c}{One-point} &\phantom{a}& \multicolumn{2}{c}{Two-point} &\phantom{a}&  \\
\cmidrule{3-4} \cmidrule{6-7}
&& C & I && C & I &&\\
\midrule
5 && 10 & 8 && 10 & 8 && 0 \\
7 && 21 & 14 && 21 & 14 && 0 \\
9 && 36 & 20 && 30 & 20 && 0 \\
11 && 37 & 24 && 30 & 23 && 1 \\
13 && 37 & 28 && 30 & 27 && 1 \\
15 && 37 & 30 && 36 & 29 && 1 \\
17 && 44 & 35 && 39 & 35 && 0 \\
19 && 46 & 39 && 39 & 37 && 2 \\
21 && 46 & 41 && 39 & 39 && 0 \\
23 && 46 & 43 && 45 & 42 && 1 \\
25 && 52 & 47 && 48 & 47 && 0 \\
27 && 54 & 50 && 48 & 48 && 0 \\
29 && 55 & 53 && 52 & 50 && 2 \\
31 && 55 & 55 && 54 & 54 && 0 \\
\bottomrule
\end{tabular}
\caption{Hermitian Curve over $\ff_{64}$. Optimal redundancy $r$ for given minimum distance $\delta$. The last column measures the improvement of the two-point improved code over the lowest redundancy in the previous columns.}
\label{Tab:64}
\end{table}

\section{Explicit Monomial Bases}

A particularly favorable feature of Hermitian curves is that one can explicitly write a monomial basis for the Riemann-Roch space of a two-point divisor. 
Fix the smooth projective plane model $ZY^q+YZ^q=X^{q+1}$. We will work with two affine charts $x=X/Z, y=Y/Z$ and $u=X/Y, v=Z/Y$, which give isomorphic affine curves $x^{q+1}=y^q+y$ and $u^{q+1}=v^q+v$. Fix points $P=P_\infty,Q=P_{(0,0)}$ (w.r.t.~the $x,y$ affine curve). Note that $x,y$ generate the ring $\bigcup_i L(iP)$ and $u,v$ generate the ring $\bigcup_i L(iQ)$. Knowing that the function $y$ has divisor $(y)=(q+1)(Q-P)$, and in particular only has zeros at $Q$, means that the ring $\bigcup_{i,j} L(iP+jQ)$ is generated by $x,y,y^{-1}$. 

\begin{lemma} 
For any $m \geq 0$ the space $L(mP)$ has a basis of the following form:
$$\{x^iy^j ~:~ 0 \leq i \leq q, iq+j(q+1) \leq m\}$$
\end{lemma}

Viewing $L(mP-aQ)$ for $a\leq q$ as a subspace of $L(mP)$, we can obtain a similar basis for $L(mP-aQ)$ by excluding the monomials that have order of vanishing at $Q$ less than $a$. For example for $q=4$, $L(10P)$ has a basis $\{1,x,y,x^2,xy,y^2\}$. Knowing that the valuation at $Q$ of $x$ is $1$ and of $y$ is $q+1$, we see that $L(10P-3Q)$ has a basis $\{y,xy,y^2\}$.

\begin{table}[ht]
\begin{tabular}{@{}lllclllclll@{}}\toprule
\multicolumn{3}{c}{Monomial}& \phantom{ab} &\multicolumn{3}{c}{Pole order at $P$} & \phantom{ab} &\multicolumn{3}{c}{Vanishing order at $Q$}\\
\midrule
$1$ & $y$  & $y^2$ &&  $0$ & $5$ & $10$ &&  $0$ & $5$ & $10$ \\
$x$ & $xy$ &       &&  $4$ & $9$ &      &&  $1$ & $6$ &     \\ 
$x^2$ &    &       &&  $8$ & &          &&  $2$ & & \\
\bottomrule
\end{tabular}
\caption{Monomials forming a basis for $L(10P)$}
\end{table}

\begin{lemma}[\cite{Park10}]\label{T:exp2pt}
Let $D = d(q+1)P - aP - bQ,$ for $d \in \ZZ,$ and for $0 \leq a,b \leq q.$ The space $L(D)$ has a basis given by the monomials $x^iy^j$ where:
\begin{enumerate}
\item $0 \leq i \leq q,$ $0 \leq j,$ and $i+j \leq d.$
\item $a \leq i \text{ for } i+j = d.$
\item $b \leq i \text{ for } j = 0.$
\end{enumerate}
\end{lemma}

Another set of explicit bases for Riemann-Roch spaces on the Hermitian curve appears in \cite{MahMatPir05}.
However, those bases are not monomial.
We illustrate the use of the explicit bases for the Hermitian curve over $\ff_{64}$ and for codes with designed distance $19$ (the line
$\delta = 19$ in Table~\ref{Tab:64}).
The classical one-point code $C_L(D,73P)^\perp$ has redundancy $r=46$ but this can be improved to $r=39$ by removing seven of the checks
(Table~\ref{Tab:example-onep}). The classical two-point code $C_L(D,74P-8Q)^\perp$ has redundancy $r=39$, which improves to $r=37$ 
after removing two of the checks (Table~\ref{Tab:example-twop}).

\begin{table}[ht]
\[\begin{array}{@{}ccccccccc@{}}\toprule
\multicolumn{9}{c}{\mbox{Monomial Checks for $C_L(D,73P)^\perp$, $\delta \geq 19$, $r=46$}}\\
\midrule
 1 & y & y^2 & y^3 & y^4 & y^5 & y^6 & y^7 & y^8 \\
 x & xy & xy^2 & xy^3 & xy^4 & xy^5 & xy^6 & xy^7 &\\
 x^2 & x^2y & x^2y^2 & x^2y^3 & x^2y^4 & x^2y^5 & (x^2y^6) &&\\
 x^3 & x^3y & x^3y^2 & x^3y^3 & (x^3y^4) & (x^3y^5) &&&\\
 x^4 & x^4y & x^4y^2 & (x^4y^3) & (x^4y^4) &&&&\\
 x^5 & x^5y & x^5y^2 & (x^5y^3 )&&&&&\\
 x^6 & x^6y & (x^6y^2) &&&&&&\\
 x^7 & x^7y &&&&&&&\\
 x^8 & x^8y &&&&&&&\\
\bottomrule
\end{array}\]
\caption{Removing the checks in parentheses reduces the redundancy to $r=39$ but preserves the distance $\delta \geq 19$.}
\label{Tab:example-onep}
\end{table}

The Feng-Rao lower bound for the weights in a dual coset can be seen explicitly by selecting the monomial that corresponds to the coset and then counting the number of monomials in the diagram that divide it \cite{FenRao93, HoeLinPel98, Bla08}. For example, in Table \ref{Tab:example-onep} the monomials $x^3y^4$ and $x^4y^3$ determine a $4 \times 5$ rectangle, which gives minimum weight $20$. Excluding them decreases the redundancy by $2$, while the words that are added to the code have weight at least $20$, preserving the designed distance at $19$. \\

Consider the difference $C_L(D,73P)^\perp \backslash C_L(D,74P)^\perp$, which amounts to adding $x^7y^2$ in Table \ref{Tab:example-onep}. The weight of a word in the 
difference is at least $3 \cdot 8=24$. Repeatedly adding $P$, we obtain a filtration $C_L(D,73P)^\perp \supset C_L(D,74P)^\perp \supset \cdots$. For each coset in the filtration, we find that the minimum weight is at least $19$ (in fact it results in a stronger bound $\delta \geq 24$). 

\begin{table}[ht]
  \[
\begin{array}{@{}ccccccccc@{}}\toprule
\multicolumn{9}{c}{\mbox{Monomial Checks for $C_L(D,74P-8Q)^\perp$, $\delta \geq 19$, $r=39$}}\\
\midrule
& y & y^2 & y^3 & y^4 & y^5 & y^6 & y^7 & y^8 \\
& xy & xy^2 & xy^3 & xy^4 & xy^5 & xy^6 & xy^7 &\\
& x^2y & x^2y^2 & x^2y^3 & x^2y^4 & x^2y^5 & x^2y^6 &.&\\
& x^3y & x^3y^2 & x^3y^3 & x^3y^4 & (x^3y^5) &.&&\\
& x^4y & x^4y^2 & x^4y^3 & (x^4y^4) &.&&&\\\
& x^5y & x^5y^2 & x^5y^3 &.&&&&\\
& x^6y & x^6y^2 &.&&&&&\\
& x^7y & x^7y^2 &&&&&&\\
 x^8 & x^8y &&&&&&&\\
\bottomrule
\end{array}\]
\caption{Removing the checks in parentheses reduces the redundancy to $r=37$. Dots are cosets in the filtration.}
\label{Tab:example-twop}
\end{table}
\begin{table}[ht]
  \[
\begin{array}{@{}ccccccccc@{}}\toprule
\multicolumn{9}{c}{\mbox{Monomial Checks for $C_L(D,73Q-7P)^\perp$, $\delta \geq 19$, $r=39$}}\\
\midrule
& v & v^2 & v^3 & v^4 & v^5 & v^6 & v^7 & v^8 \\
& uv & uv^2 & uv^3 & uv^4 & uv^5 & uv^6 & uv^7 &\\
.& u^2v & u^2v^2 & u^2v^3 & u^2v^4 & u^2v^5 & u^2v^6 &&\\
.& (u^3v) & u^3v^2 & u^3v^3 & u^3v^4 & u^3v^5 &&&\\
.& (u^4v) & u^4v^2 & u^4v^3 & u^4v^4 &&&&\\
.& u^5v & u^5v^2 & u^5v^3 &&&&&\\
.& u^6v & u^6v^2 &&&&&&\\
 u^7 & u^7v &&&&&&&\\
 u^8 & u^8v &&&&&&&\\
\bottomrule
\end{array}
\]
\caption{Removing the checks in parentheses reduces the redundancy to $r=37$. Dots are cosets in the filtration.}
\label{Tab:example-twop2}
\end{table}

Table \ref{Tab:example-twop} shows the checks for the two-point code $C_L(D,74P-8Q)^\perp.$
Similar to the one-point case, the $P$-cosets corresponding to monomials $x^4y^4$ and $x^3y^5$ have minimum weights at least $20$ and can be removed. 
However, checking that the code itself has minimum distance $\geq 19$ by the $P$-filtration method fails. The step $C_L(D,79P-8Q)^\perp \supset C_L(D,80P-8Q)^\perp$ in the filtration corresponds to monomials $xy^8$ and $x^{10}$, both with pole order $80$ at $P$. The monomials that divide $xy^8$ form a two-by-eight rectangle. In addition,
there is a single monomial $x^8$ that divides $x^{10}$, resulting in the bound $2 \cdot 8 + 1 = 17$ for the minimum weight in the coset. 
The breakdown of the $P$-filtration is essentially the same phenomenon that we saw earlier in Figure \ref{Fig:inconsist}. The estimates for the $P$-cosets can be improved if we also consider filtrations with $Q$-cosets \cite{Bee07, DuuPar10}. However, $Q$-coset bounds cannot be seen immediately if we use monomials generated by $x$ and $y$. 

A solution to this obstacle is to switch the roles of $P$ and $Q$ and to consider an equivalent code for which the basis belongs to a space $L(mQ)$.
As a consequence, we work with $u, v$ monomials.  
After the five dots are filled in, the code $C_L(D,79P-8Q)^\perp$ is equivalent to $C_L(D,73Q-2P)^\perp$. 
The step $C_L(D,73Q-2P)^\perp \backslash C_L(D,74Q-2P)^\perp$ in the $Q$-filtration corresponds to a monomial $u^7v^2$.
The monomials in $L(74Q)$ that divide $u^7v^2$ form a eight-by-three rectangle; thus, the monomials in $L(74Q-2P)$ are two less and the bound is $22$.
The monomials in Table \ref{Tab:example-twop2} form a basis for $L(73Q - 2P)$.
As expected by Theorem \ref{T:exp2pt} this is just $L(73Q)$ with $1,u$ omitted. 
The desired design distance of $19$ is shown to hold by continuing further with $Q$-cosets. 
In fact, the resulting bound for the code $C_L(D,73Q-7P)^\perp$ is slightly better and it equals $21$.

\def\lfhook#1{\setbox0=\hbox{#1}{\ooalign{\hidewidth
  \lower1.5ex\hbox{'}\hidewidth\crcr\unhbox0}}}
\
\section*{Acknowledgement}
The authors would like to thank the anonymous referees for their comments.

\begin{IEEEbiographynophoto}
{Iwan} M. Duursma received his PhD in
Mathematics from the University of Eindhoven in 1993. After positions with
CNRS IML Luminy, University of Puerto Rico, Bell-Labs, AT\&T Research, and
University of Limoges, he is currently associate professor at the
University of Illinois at Urbana-Champaign.
\end{IEEEbiographynophoto}
\begin{IEEEbiographynophoto}
  {Radoslav} Kirov received his PhD in Mathematics from the University of Illinois at Urbana-Champaign, under the supervision of Iwan Duursma. Currently, he is a research fellow at Nanyang Technological University, Singapore.
\end{IEEEbiographynophoto}
\end{document}